\documentclass[floatfix, a4paper, amsfonts, amssymb, amsmath, reprint, showkeys, nofootinbib, twoside, superscriptaddress, aps, prb]{revtex4-2}

\usepackage{graphicx} 
\usepackage{physics}
\usepackage{mathtools}
\usepackage{colonequals}
\usepackage[caption=false]{subfig}
\usepackage{nicefrac}
\usepackage{siunitx} 
\usepackage{color, colortbl} 
\usepackage{xcolor, colortbl} 
\usepackage{wasysym}
\usepackage{dsfont}

\usepackage{multirow}
\usepackage[normalem]{ulem} 
\usepackage{float}
\usepackage{xr}
\usepackage{array}
\usepackage{hyperref}
\usepackage[T1]{fontenc}
\usepackage{notes2bib}
\bibnotesetup{
  note-name = ,
  use-sort-key = false
}
     
\newcommand{\kk}{{\vb k}}
\newcommand{\EE}{{\vb E}}

\newcommand{\pcm}{$\si{\per\centi\meter}$}
\newcommand{\um}{$\si{\micro\meter}$}
\newcommand{\pum}{$\si{\micro\meter}^{-1}$}

\begin{document}
\title{Dispersion of backward-propagating waves in a surface defect \\on a 3D photonic band gap crystal} 

\author{Timon J. Vreman}
\affiliation{Complex Photonic Systems (COPS), MESA\texttt{+} Institute, University of Twente, 7500 AE Enschede, The Netherlands}
\author{Melissa J. Goodwin}
\affiliation{Complex Photonic Systems (COPS), MESA\texttt{+} Institute, University of Twente, 7500 AE Enschede, The Netherlands} 

\author{Lars J. Corbijn van Willenswaard}
\affiliation{Complex Photonic Systems (COPS), MESA\texttt{+} Institute, University of Twente, 7500 AE Enschede, The Netherlands}
\affiliation{Mathematics of Computational Science (MACS), MESA\texttt{+} Institute, University of Twente, 7500 AE Enschede, The Netherlands}

\author{William L. Barnes}
\affiliation{Department of Physics and Astronomy, University of Exeter, Exeter EX4 4QL, United Kingdom}
\affiliation{Complex Photonic Systems (COPS), MESA\texttt{+} Institute, University of Twente, 7500 AE Enschede, The Netherlands}

\author{Ad Lagendijk}
\affiliation{Complex Photonic Systems (COPS), MESA\texttt{+} Institute, University of Twente, 7500 AE Enschede, The Netherlands}

\author{Willem L. Vos}
\affiliation{Complex Photonic Systems (COPS), MESA\texttt{+} Institute, University of Twente, 7500 AE Enschede, The Netherlands}
\email{w.l.vos@utwente.nl}

\begin{abstract}
We experimentally study the dispersion relation of waves in a two-dimensional (2D) defect layer with periodic nanopores that sits on a three-dimensional (3D) photonic band gap crystal made from silicon by CMOS-compatible methods. 
The nanostructures are probed by momentum-resolved broadband near-infrared imaging of p-polarized reflected light that is collected inside the light cone as a function of off-axis wave vectors. 
We identify surface defect modes at frequencies inside the band gap with a narrow relative linewidth ($\Delta\omega/\omega = 0.028$), which are absent in defect-free 3D crystals.
We calculate the dispersion of modes with relevant mode symmetries using a plane-wave-expansion supercell method with no free parameters. 
The calculated dispersion matches very well with the measured data.
The dispersion is negative in one of the off-axis directions, corresponding to backward-propagating waves where the phase velocity and the group velocity point in opposite directions, as confirmed by finite-difference time-domain simulations. 
We also present an analytic model of a 2D grating sandwiched between vacuum and a negative real $\epsilon' < 0$ that mimics the 3D photonic band gap. 
The model's dispersion agrees with the experiments and with the fuller theory and shows that the backward propagation is caused by the surface grating.
We discuss possible applications, including a device that senses the output direction of photons emitted by quantum emitters in response to their frequency. 
\end{abstract}

\maketitle

\section{Introduction}\label{sec:introduction}
There is a worldwide interest in completely controlling the propagation and emission of photons, a major outstanding goal of the field of nanophotonics~\cite{Sakoda2005Springer, Joannopoulos2008book, Lourtioz2008book, Noginov2009book, Novotny2012book, Ghulinyan2015book}. 
An intriguing kind of control is the opportunity to have photons propagate preferentially in a thin quasi-two-dimensional (2D) sheet in space, where propagation in the 3$^\mathrm{rd}$ dimension is exponentially attenuated or otherwise somehow impeded. 
Such peculiar ``flattened'' propagation is well-known on a single interface between a dielectric (with $\epsilon_{d}' > 0$) and a metal (with $\epsilon_{m}' < 0$),\footnote{The dielectric constant $\epsilon = \epsilon' + i \epsilon''$ is written as a sum of a real part $\epsilon' = \rm{Re}\{\it{\epsilon}\}$ and an imaginary part $\epsilon'' = \rm{Im}\{\it{\epsilon}\}$.} where surface plasmon polaritons occur~\cite{barnes2003nature, maier2007springer, Saleh2019Wiley}; and the double-interfaced counterpart~\cite{Koppens_NL_2011_11_337}. 
A second way to confine light to 2D occurs on the surface of a 3D photonic band gap crystal,\footnote{Photonic crystals are nanostructures with spatially periodic variations of the dielectric constant, with length scales $a$ of the order of the wavelength of light: $a \simeq \lambda$. } where surface states occur~\cite{Joannopoulos2008book,takayama2017CondensMatter} that have been observed in pioneering studies by Ishizaki \textit{et al.}~\cite{Ishizaki2009nature, Ishizaki2013OptExpr}. 
A third approach is to confine light to a thin single-mode slab with a high dielectric constant by total internal reflection, as is widely pursued in 2D slab photonic crystals that have periodic arrays of pores in the slab~\cite{krauss1996nature,johnson1999PRB}. 
In all these platforms, light has wave vectors exceeding those in free space ($\abs{\kk_{}} \equiv k_0 > \abs{\kk_{\rm{vac}}}$), i.e. outside the light cone so that special incoupling/outcoupling methods are needed to overcome the momentum mismatch, such as gratings or prisms. 

\begin{figure}[t]
    \centering
    \includegraphics[width=\linewidth]{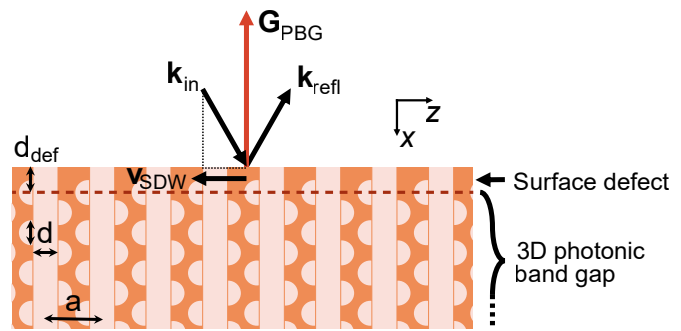}
    \caption{Schematic cross-section showing the excitation of waves in a planar surface defect (thickness $d_{\rm{def}}$, pitch $a$, pore diameter $d$) on the surface of a 3D photonic band gap crystal (orange). 
    The surface defect wave (SDW) has a group velocity $\vb{v}_{\rm{SDW}}$ pointing in the opposite direction as the momentum of the incident wave parallel to the surface along z: $k_{\rm{in,z}}$.
    Part of the incident light is reflected to $\kk{}_{\rm{refl}} \equiv \kk{}$ due to Bragg scattering by reciprocal lattice vector $\vb{G}_{\rm{PBG}}$ where $\kk{} = \kk{}_{\rm{in}}$ + $\vb{G}_{\rm{PBG}}$. 
    }
    \label{fig:illustrationSurfWave}
\end{figure}

Here, we study waves that propagate in a thin dielectric photonic layer on the surface of a 3D photonic band gap crystal. 
Such a thin layer may also be viewed as a 2D surface defect on a 3D photonic band gap crystal, which was recently proposed as an effective absorber~\cite{Sharma2021OptEx}. 
Since a 3D photonic band gap may be viewed as an effective medium with a negative real dielectric constant ($\epsilon' < 0$), the light in the dielectric layer is exponentially attenuated into the lower half space. 
In our case, the thin dielectric layer has a periodic array of pores, as illustrated in the schematic in Fig~\ref{fig:illustrationSurfWave}. 
Thus, our surface defect layer is structured, which may be conceived as a 2D slab photonic crystal, which plays the role of an incoupling or outcoupling grating for the light propagating inside the layer. 
Thus, our system operates inside the light cone, and the light in the surface defect layer is not strictly confined from free space~\cite{Saleh2019Wiley}. 
We 
probe the dispersion of the surface defect light using momentum-resolved imaging~\cite{zhang2018PRL, chen2019AcsPhot, cueff2024Nanophotonics} of the reflected light. 
Remarkably, the reflected light reveals narrow minima, indicating good confinement from free space, and the minima are deep, indicating an efficient incoupling ($\sim{90}$\%) from free space. 
These properties are distinct from those of the approaches listed above that -- in the visible part of the spectrum -- often possess rather broad minima when excited within the light cone. 
In contrast to metal-based plasmonics, our system has a vanishing $\epsilon''$, so we expect minimal non-radiative damping due to absorption. 
A second remarkable feature of our hybrid 2D+3D system is that the waves reveal backward propagation~\cite{foteinopoulou2007prb, foteinopoulou2007apl, vinogradov2010surface, takayama2017CondensMatter, hu2018optLett,yilmaz2018OptLett, gonzalez2020OptLett}, 
where the group velocity $\vb{v}_{\rm{SDW}}$ is opposite to $k_{\rm{in,z}}$, see Fig.~\ref{fig:illustrationSurfWave}, as confirmed by simulations. 
The backward propagation results from the scattering of incident light by the periodically structured surface layer, as borne out by an analytic model that involves a surface grating on top of a negative $\epsilon$ band gap medium. 

To interpret our experiments, we performed plane-wave expansion calculations with supercells and find that the surface defect allows for predominantly p-polarized light that propagates inside the added dielectric material at frequencies matching the photonic band gap. 
The calculated dispersion agrees very well with the measured dispersion, which is exhilarating in view of no free parameters. 
We conclude our paper by discussing a few potential device applications.

\begin{figure}[t]
    \centering
    \includegraphics[width=\columnwidth]{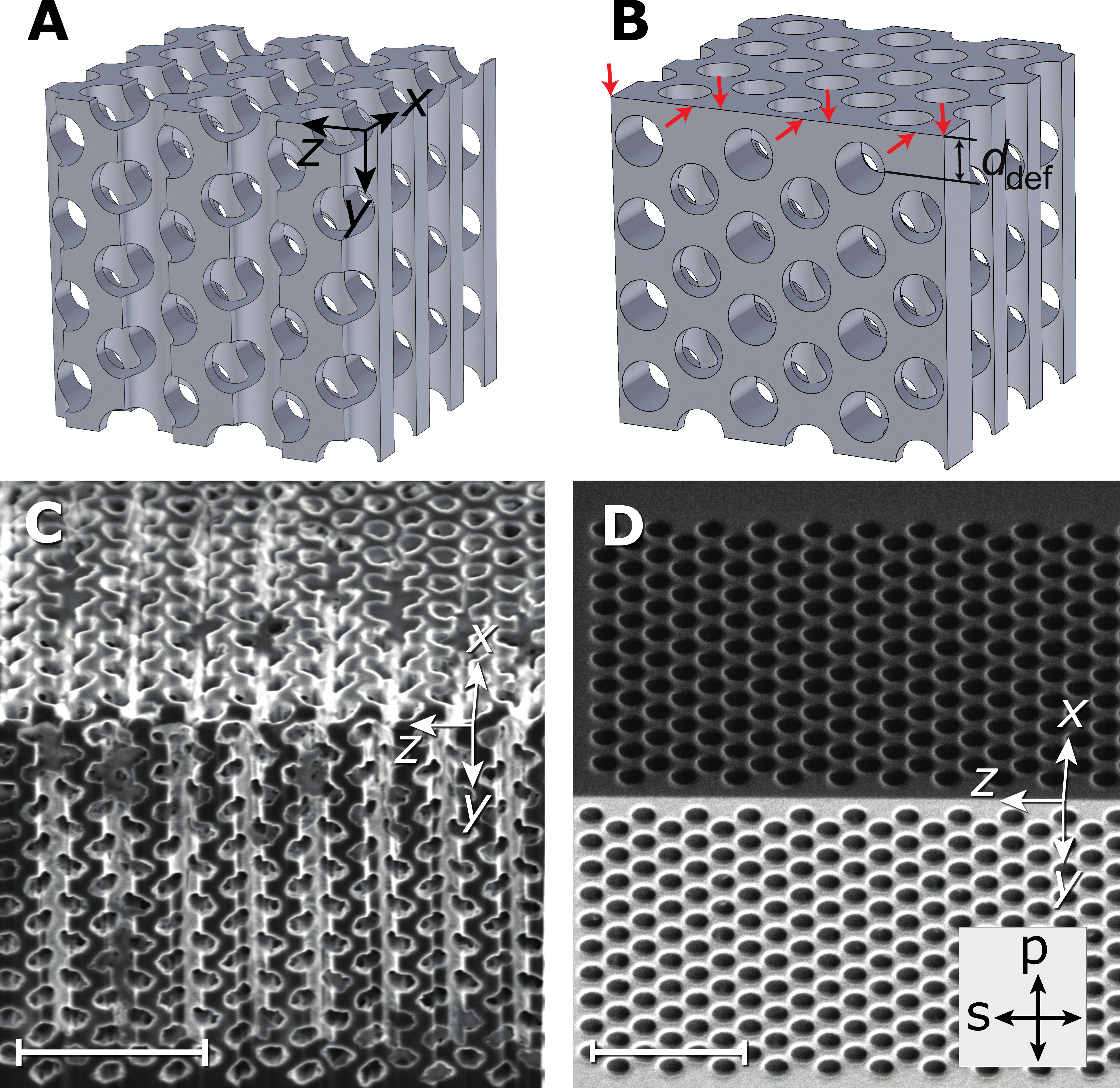}
    \caption{
    3D photonic band gap crystals (A,C) without and (B,D) with surface defect. (A) Model of a periodic crystal which is periodic up to the crystal-to-air interface.
    (B) Model of the same structure, missing half-pores at the red arrows, creating a surface defect.
    (C) SEM image of the real photonic crystal after the surface was sliced off, similar to (A).
    (D) SEM image of the mask of the same structure as (C) before the surface was sliced off, therefore having the surface defect, similar to (B).
    Scale bars shown are 2~$\si{\micro\meter}$ wide.
    }
    \label{fig:sample3d}
\end{figure}

\section{Experimental section}\label{sec:experimental}
\subsection{Samples}\label{sec:samples}
Our 3D photonic band gap crystals have the inverse woodpile crystal structure~\cite{Ho1994SSC} and are made of silicon by CMOS-compatible means. 
Inverse woodpile crystals have a broad 3D photonic band gap because of their cubic diamond-like structure~\cite{maldovan2004NatureMater}. 
The inverse woodpile structure consists of two identical arrays of pores with radius $R$ running in the perpendicular $x$- and $y$-directions.
Each array of pores has a centered-rectangular structure with lattice constants $a$ and $c$ in a ratio $a/c = \sqrt{2}$ to ensure a cubic crystal structure. 
Here, we use $a = 680$~nm and a pore diameter $d = 300$~nm to produce a photonic band gap in the near infrared and telecom ranges~\cite{Huisman2011PRB,vandenBroek2012AFM,Devashish2017PRB,Adhikary2020OptEx}.
The pore arrays on the $yz$-surface have an offset $\Delta z = a/4 = 170$~nm in the $z$-direction compared to the $xz$-surface, see Fig~\ref{fig:sample3d}.
For more detailed information on the unit cell, see Appendix~\ref{sec:latticereal}. 

The 3D silicon photonic crystals are made by etching deep cylindrical pores on the edge of a silicon beam in the $x$- and $y$-directions. 
We made the nanostructures using techniques we have described before~\cite{ vandenBroek2012AFM,Grishina2015Nanotech,goodwin2023Nanotech}. 
In our standard mask-making procedure~\cite{Grishina2015Nanotech} we came to realize that we introduce a peculiar surface termination: from half of the pores at the $xz$- and $yz$-surfaces, the lower half of the pore is missing compared to a perfectly periodic structure that is terminated halfway through the pore, as in \textit{e.g.}, Refs.~\cite{Devashish2017PRB,Mavidis2020PRB}.
Therefore, our fabricated structures have from the outset a surface defect, see Fig.~\ref{fig:sample3d}(B,D). 
The thickness of the surface defect in Fig.~\ref{fig:sample3d}(D) is $d_{\rm{def}} = 320$~$\pm 10$~nm, where $d_{\rm{def}}$ is defined as the distance from the surface to the centers of the pores closest to the surface.
After experiments on the structure with a surface defect are completed, we carefully removed part of the surface of the photonic crystal with a focused ion beam (FIB) such that the structure is periodic up to the surface, see Fig.~\ref{fig:sample3d}(A,C).
Similar measurements were repeated on a second photonic crystal both with and without surface defect, see Appendix~\ref{sec:app_measuredsamples}, and were found to reproduce very well. 

\subsection{Optical setup}\label{sec:setup}

\begin{figure}[t]
    \centering
    \includegraphics[width=\columnwidth]{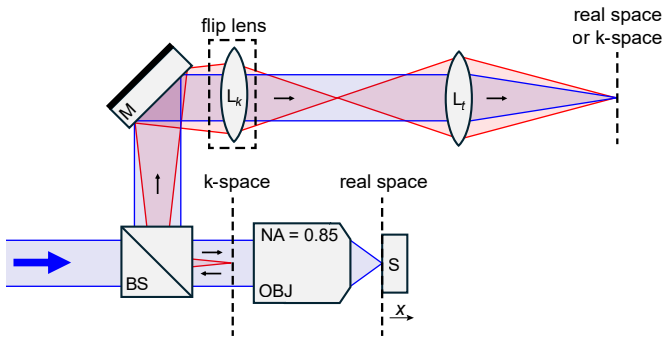}
    \caption{
    Optical setup to measure spectrally-resolved reflectivity both in real space and in wave vector space at the detector plane (vertical dashed line, top right). 
    The incident beam has a tunable wavelength between 1000 and 1700~nm.
    The blue beam path pertains to real-space imaging of the sample. 
    The red beam path pertains to imaging in wave vector space (k-space). 
    Components shown: a flip lens ($\rm{L}_{\it{k}}$), a beam splitter (BS), an objective (OBJ), a sample (S), a mirror (M), and a tube lens ($\rm{L}_{\it{t}}$). 
    }
    \label{fig:setup}
\end{figure}

Although band gaps and surface defects are typically studied by scanning the angle of incident light~\cite{Vos2000PLA, noda2000science, dedood2003prb, qi2004nature, man2005Nature, Ishizaki2009nature, takahashi2009NatMat, Suzuki2011OptExpr, Ishizaki2013OptExpr, chen2019AcsPhot}, we chose to scan the frequency while directly imaging the reflectivity as a function of off-axis momentum, a method called momentum-resolved imaging or Fourier imaging~\cite{chen2019AcsPhot,le_thomas2007JOptSocAmB, chen2019AcsPhot, Zhang2021ScienceB, zhang2022prb, cueff2024Nanophotonics}.
Figure~\ref{fig:setup} schematically shows our setup to collect spectrally-resolved reflectivity in both real and wave vector space. 
Our setup is based upon that used in Refs.~\cite{Adhikary2020OptEx, Uppu2021PRL}. 
Briefly, linearly polarized light from a supercontinuum white light source (Fianium SC-400-2) is spectrally filtered by a monochromator with spectral resolution $\Delta \lambda$ = 0.6~nm (Oriel MS257), collimated, and focused ($\diameter < 2$~$\um{}$) on the samples with an objective (Olympus LCPLN100XIR) with a numerical aperture $\rm{NA} = 0.85$. 
Reflected light is collected in backscatter geometry by the same objective and directed via a beam splitter through a polarizer to an InGaAs camera (Photonic Sciences). 
Waves traveling in the $x$-direction are labeled as s-polarized (p-polarized) if the $\EE{}$-field is parallel to the $z$-axis ($y$-axis). 
The half-wave plate and the polarizer are set to both s-polarized or both p-polarized. 
To collect spectral information in $\kk$-space, light collected with the objective is collimated, and the back focal plane is imaged by two lenses (flip lens and tube lens, 300 and 500~mm, respectively) onto a camera.
A frequency scan in $\kk$-space of 160 data points takes 5 minutes per sample per polarization. 

\subsection{Experimental details}\label{sec:dataanalysis}
The reflectivity is normalized relative to the reflectivity measured with a gold mirror and assuming $R_{\rm{Au}} = 96 \%$ reflectivity. 
As the component of the wave vector parallel to the air-sample interface $\kk_{\parallel}$ is conserved~\cite{born1999cambridge}, we plot $k_y$ and $k_z$ on the axes instead of angle. 
The maximum measurable $k_y$ and $k_z$, also known as the light line, is equal to the NA times the magnitude of the wave vector in air $k_0$, \textit{i.e.}, $k_{\parallel,\rm{max}}$ = $(\rm{NA})k_0 $.

Each measurement run yields a 3D dataset - two dimensions are spatial ($k_y$ and $k_z$) and one is the frequency ($\tilde{\nu}$).
Videos in the Supplementary Material show whole datasets, also for a second photonic crystal with the same design parameters~\cite{supMat_c2_noDef}.
For simplicity, here we will focus on either one frequency (2 spatial dimensions), or we will set $k_y = 0$ or $k_z = 0$ (1 spatial and 1 frequency dimension).

\subsection{Photonic band structure calculations}\label{sec:theorybands}
\begin{figure}[t]
    \centering
    \includegraphics[width=0.5\columnwidth]{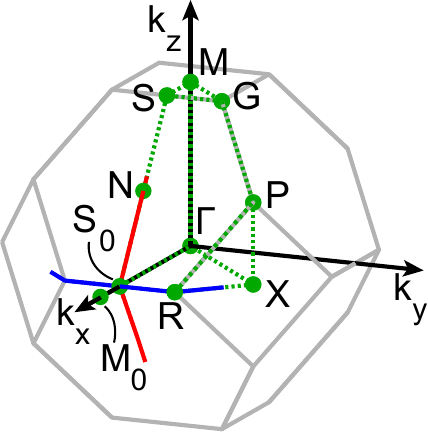}
    \caption{
    (Green, dotted) Irreducible Brillouin zone for a 3D photonic band gap crystal with the inverse-woodpile structure. 
    (Gray) First Brillouin zone. 
    (Blue/red) Trajectories in k-space along which we computed bands to compare to the experiment. 
    The trajectories extend until $k_y/|\kk{}|$ = $k_z/|\kk{}|$ = NA = 0.85 for $\tilde{\nu} = 10^4$~$\si{\centi\meter}^{-1}$.
    }
    \label{fig:ibz_fbz}
\end{figure}

We determine the edges of the 3D photonic band gap of the crystal without defect as a function of off-axis momentum for p-polarized light as follows: 
The band structures for Bloch modes are calculated for wave vectors on the boundary of the first Brillouin zone, see Fig.~\ref{fig:ibz_fbz}, as those wave vectors typically determine the frequency edges of the band gap. 
The $\Gamma \rm{S_0}$-direction corresponds to waves incident from the normal $x$-direction.
By setting $k_y = 0$ or $k_z = 0$, two paths are defined along the first Brillouin zone boundary, see the red and blue lines in Fig.~\ref{fig:ibz_fbz}, respectively. 
Modes with $\kk{}$-vectors along the red and blue lines are calculated using the MIT Photonic Bands (MPB) open-source software~\cite{johnson2001OptExp}, using a pore diameter $d$ = 300~nm, and lattice parameter $a$ = 680~nm (reduced pore radius $r/a = 0.22$).
The first Brillouin zone of the inverse-woodpile structure is described further in Appendix~\ref{sec:latticereci}.

Bloch modes in a photonic crystal have symmetry properties~\cite{Sakoda2005Springer}, which we also calculate using the MPB software.
The symmetry properties of modes with wave vectors along the $\Gamma$$\rm{S_0}$ path are categorized into four types, namely {$\Sigma_1$} to {$\Sigma_4$}~\cite{aroyo2006bilbao2, elcoro2017ApplCryst}, as elaborated on in Appendix~\ref{sec:symalongpore}. 
These wave vectors correspond to plane waves incident from the $x$-direction.
Due to the symmetry properties of polarized plane waves, s-polarized and p-polarized waves excite only modes from {$\Sigma_3$} and {$\Sigma_4$}, respectively. 

At the photonic crystal-to-air interface, the parallel momentum and frequency (and wavenumber $\tilde{\nu}$) are conserved and can be directly plotted in wavenumber versus off-axis momentum graphs. 
In these band structures, we only show the modes with the symmetry corresponding to the polarization of that measurement. 
We also do this for the off-axis modes, as the symmetry properties are expected to still dominate.

\subsection{Angle of reflection off photonic crystals}\label{sec:probe}
For a planar mirror the angle of incidence equals the angle of reflection due to the conservation of in-plane momentum; 
the incoming parallel momentum $\kk_{\rm{in},\parallel}(\tilde{\nu})$ equals the outgoing parallel momentum $\kk_{\parallel}(\tilde{\nu})$ at a certain wavenumber $\tilde{\nu}$.~\footnote{The wavenumber $\tilde{\nu} = \omega / 2 \pi c_{0}$ is in spectroscopic tradition gauged in $\mathrm{cm}^{-1}$ that is equivalent to the older unit kayser~\cite{wikiWavenumber}.}
Additionally, a mirror reflects all angles (almost) equally.

The interface between air and a 3D photonic band gap crystal is more complex~\cite{Notomi2000prb}. 
If a wave with $\kk_{\rm{in},\parallel}(\tilde{\nu})$ excites a Bloch mode, the reflection of that wave is expected to be low.
And if such a wave does not excite a Bloch mode, Bragg diffraction scatters $\kk_{\rm{in}}$ to $\kk{}$, such that
\begin{equation}
    \kk - \kk_{\rm{in}} = \mathbf{G},
\label{eq:laue}
\end{equation}
where $\mathbf{G}$ is a reciprocal lattice vector~\cite{Ashcroft1976book} or a sum of reciprocal lattice vectors in the case of multiple Bragg diffraction~\cite{tajiri2020prb}. 
Therefore, measuring (low) high reflectivity at a certain $\kk_{\parallel}(\tilde{\nu})$ does not necessarily mean that $\kk_{\rm{in},\parallel}(\tilde{\nu})$ with the same momentum is exciting Bloch modes (well) poorly. 
Moreover, it is possible to observe more than 100\% reflectivity at certain $\kk_{\parallel}$ due to redirection.

To link the measured reflectivity with the calculated band structures, we assume that the reflectivity is dominated by a reciprocal lattice vector $\mathbf{G}$ pointing normal to the surface, \textit{i.e.}, $\mathbf{G}_{\rm{dom}}$ = $G_{\perp}\mathbf{\hat{x}}$.
The lower the frequency of light, or dimensionality of the structure (such as our quasi-2D surface defect), the better this assumption is. 
Using Eq.~\ref{eq:laue}, we then find $\kk_{\rm{in},\parallel}$ = $\kk_{\parallel}$, just like a mirror (see Fig.~\ref{fig:illustrationSurfWave}).
Note that when this assumption does not hold, the measured reflectivity and band calculations are still correct, but they will likely not match well.

The blue path between $\rm{S_0}$ and $\rm{R}$ in Fig.~\ref{fig:ibz_fbz} traverses the intersection of two Bragg planes, and therefore multiple Bragg diffraction occurs~\cite{vanDriel2000PRB, tajiri2020prb}; Point $R$ involves the intersection of three Bragg planes.
In contrast, the red path $\rm{S_0}$ to $N$ of Fig.~\ref{fig:ibz_fbz} runs along a \textit{single} Bragg plane.
Therefore, a different behavior is expected between the red and blue paths, such as a different $\rm{\mathbf{G}}_{\rm dom}$.

\subsection{FDTD simulation}\label{sec:expMeepSims}
The excitation of the surface defect mode is simulated on an ideal 3D photonic crystal of 6$c\cross$5$c\cross$40$a$ with a surface defect with thickness $d_{\rm{def}} = 320$~nm.
In finite-difference time-domain simulations (FDTD) using open-source package Meep~\cite{oskooi2010elsevier}, a p-polarized quasi-monochromatic plane wave with $k_z =$ \texttt{+}$2.0$~\pum{} and $\tilde{\nu}_{\rm{SDW}} = 6500$~\pcm{} is incident on the photonic crystal, which corresponds to the $\kk{}_{\rm{in}}$ that excites the surface defect mode.
The simulation domain is surrounded by 1.5~\um{} of perfectly matched layer (PML) to absorb outgoing waves.
Cross-sections of the electric field in the $xz$-plane are extracted from the simulation to show the propagation of the guided wave. 

\subsection{Supercell calculation of the surface defect mode}\label{sec:expSupercell}
To calculate the dispersion of the surface defect mode, we apply the supercell method~\cite{meade1991prb}. 
For this calculation, the orthorhombic unit cell is used, as previously used~\cite{Hillebrand2003JAP,Woldering2009JAP,Devashish2017PRB,Adhikary2020OptEx,Huisman2011PRB}, with the pores at the same position as in Appendix~\ref{sec:latticereal}.
Six unit cells of 3D photonic crystal are alternated with six unit cells of air in the $x$-direction, with a resolution of 192$\cross$16$\cross$23. 
Of the outer two unit cells, one has a surface defect with thickness $d_{\rm{def}}$ = 320~nm, and the other has proper surface termination halfway through the pores.

Modes are calculated at $\kk{}$-vectors with either $k_y$ = 0, or $k_z$ = 0 while keeping $k_x$ constant. 
Next to the surface defect mode and bulk crystal modes, the supercell method also finds modes in the air unit cells, which interact with and obscure the surface mode.
The confinement of the surface defect modes near the surface and the dominant polarization of the defect mode are used to isolate the surface defect mode, as elaborated on in Appendix~\ref{sec:app_supercell}.

\section{Results}\label{sec:results}

\subsection{Reflectivity at a single frequency}\label{sec:res_singleFreq}
\begin{figure}[t]
    \centering
    \includegraphics[width=\columnwidth]{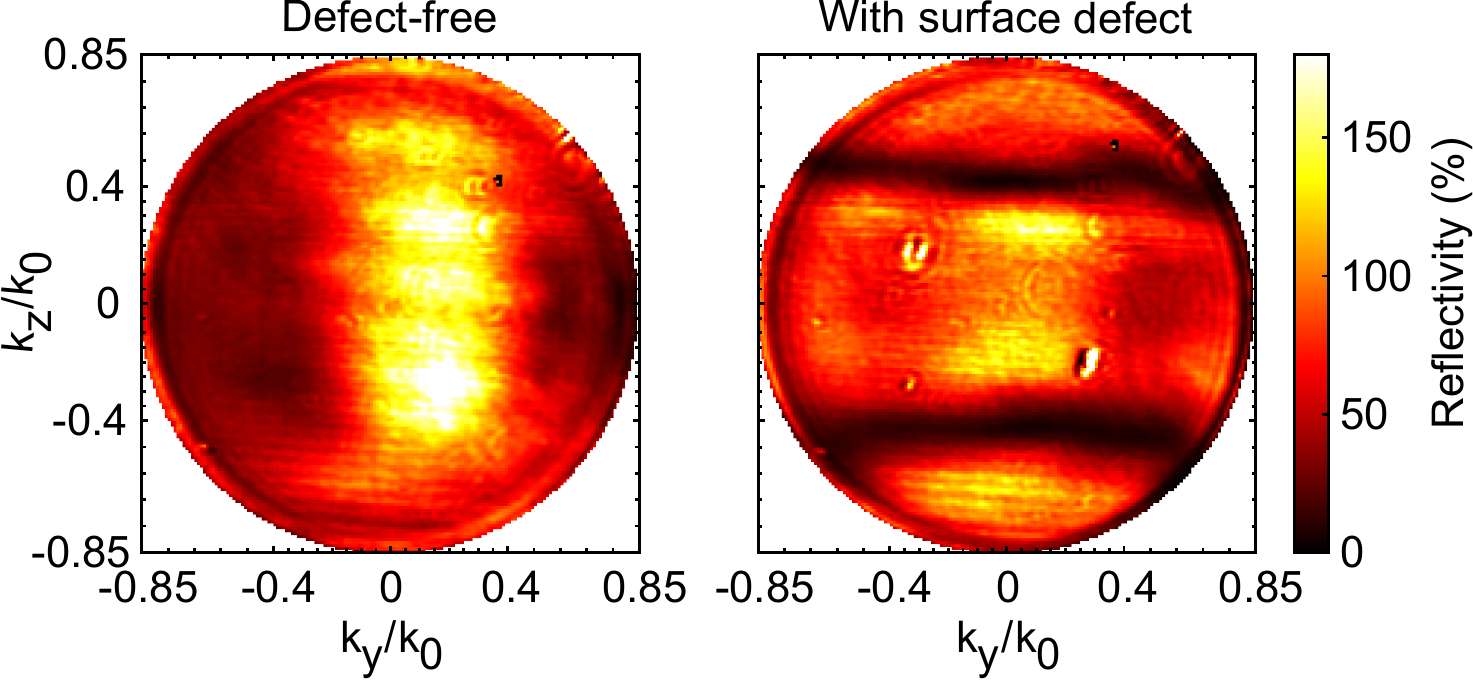}
    \caption{
    Momentum-resolved image of the reflectivity of the photonic crystal (left) without and (right) with defect at $\tilde{\nu} = 6850$~\pcm{} ($\lambda = 1460$~nm; $a/\lambda = 0.4657$) for p-polarized light.
    We observe high reflectivity from the band gap with strong line-like troughs from the defect mode.
    The NA of the objective determines the maximum relative off-momentum, 0.85.
    }
    \label{fig:kspaceSingleFreq}
\end{figure}

The momentum-resolved reflectivity of p-polarized light is shown in Fig.~\ref{fig:kspaceSingleFreq} at a frequency ($\tilde{\nu}$ = 6850~\pcm{}) deep inside the band gap.
Without the surface defect, the reflectivity is high, as expected from a complete 3D band gap.
In the $k_z$-direction, the reflectivity is high everywhere; 
in the $k_y$-direction, the reflectivity is greater than 100\% near $k_y=0$, and lower at $\abs{k_y} > 0$, which we attribute to the redirection of the incident light from higher angles being reflected to $k_y = 0$.
The reflectivity is centered slightly towards positive $k_y$, as the surface removal process by the FIB results in a small angular offset.
Other than that, the reflectivity is symmetric around both $k_y = 0$ and $k_z = 0$, as expected from a symmetric structure.

For the crystal with the surface defect, we also observe high reflectivity, and the reflectivity is mostly symmetric around $k_y = 0$ and $k_z = 0$.
Remarkably, we observe two deep troughs, tending to a minimum reflectivity as low as $\sim$10\%, resulting from the surface defect. 
The troughs are shaped like two lines: they extend along all probed $k_y$, and they are sharp along $k_z$ appearing at $\abs{k_z} = 0.4 k_0$.
As a function of increasing frequency, the troughs from Fig.~\ref{fig:kspaceSingleFreq} approach each other, as is observed in the video of the whole momentum-resolved data set in the Supplementary Material~\cite{supMat_c2_noDef}.

Since the minima are absent in the reflectivity acquired from the sample without a defect, we conclude that they correspond to new states from the surface defect.
The minima correspond to the excitation of waves that travel between the band gap and air, as illustrated in Fig.~\ref{fig:illustrationSurfWave}.
Because the wave is excited with light inside the light cone, radiation into the air is expected.
We further investigate and discuss the radiation in Secs.~\ref{sec:res_simulations} and~\ref{sec:discussion}.

\subsection{Frequency-resolved reflectivity of 3D band gap}\label{sec:res_noSurfDef}
\begin{figure}[t]
    \centering
    \includegraphics[width=\columnwidth]{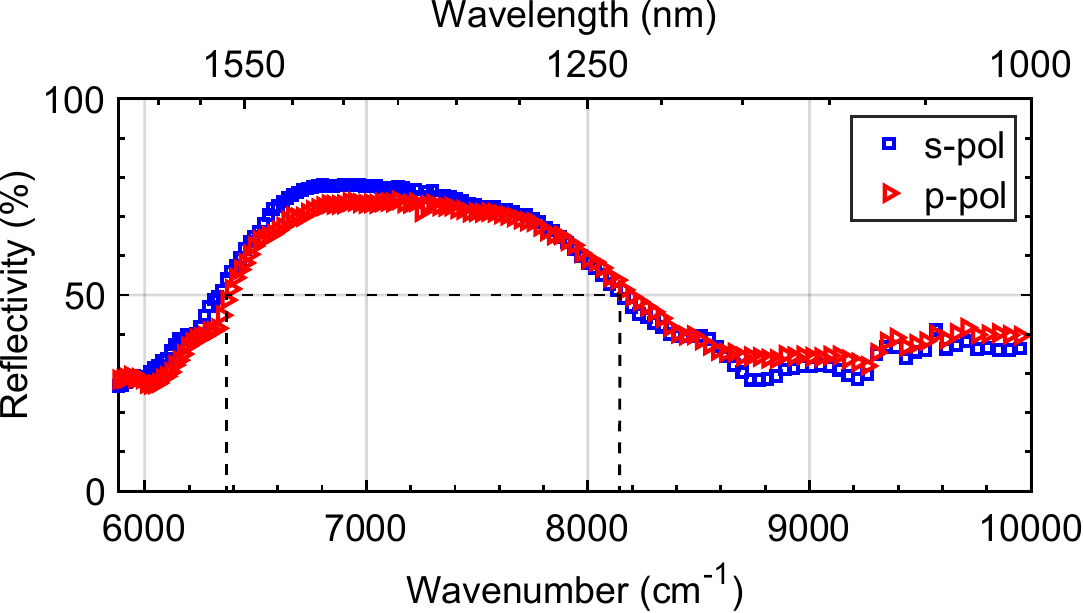}
    \caption{
    Total reflectivity of the 3D photonic band gap crystal without surface defect, as shown in Fig.~\ref{fig:sample3d}(C). 
    Blue data are for s-polarization, red data for p-polarization. 
    Horizontal dashed line indicates the half height and vertical dashed lines indicate the full width at half maximum. Wavenumber $\tilde{\nu} = \omega/2\pi c_0$.
    }
    \label{fig:realreflSpectrum}
\end{figure}

By integrating the reflectivity over momentum space, we obtain real-space reflectivity spectra, similar to previous studies in our group~\cite{Huisman2011PRB,Adhikary2020OptEx}, and elsewhere~\cite{Schilling2005apl, garcia2007AdvMat, takahashi2009NatMat}.
The total reflectivity spectra measured on the structure without defect in Fig.~\ref{fig:realreflSpectrum} reveal broad and intense peaks centered at $\tilde{\nu}_c$ = 7200~\pcm{} for both polarizations with maxima up to 70-80\%, confirming the high quality of our crystal structures.
The peaks are broad with widths (full width at half maximum) of about $\Delta\tilde{\nu}_c$ = 1800~\pcm{}, corresponding to a relative bandwidth of $\Delta\tilde{\nu}_c$/$\tilde{\nu}_c$ = 25\% and thus confirming the expected high photonic strength of these high-index-contrast silicon-air nanostructures~\cite{Huisman2011PRB, Adhikary2020OptEx}.
The reflectivity spectra are smooth for both polarizations.
The band gap is expected to extend until 8200~\pcm{} (see the band structure in Fig.~\ref{fig:banddiagram}), but we observe that the reflectivity starts to decrease at 7800~\pcm{}, which we attribute to minor manufacturing defects, such as non-cylindrically shaped pores.

\begin{figure}[t]
    \centering
    \includegraphics[width=\columnwidth]{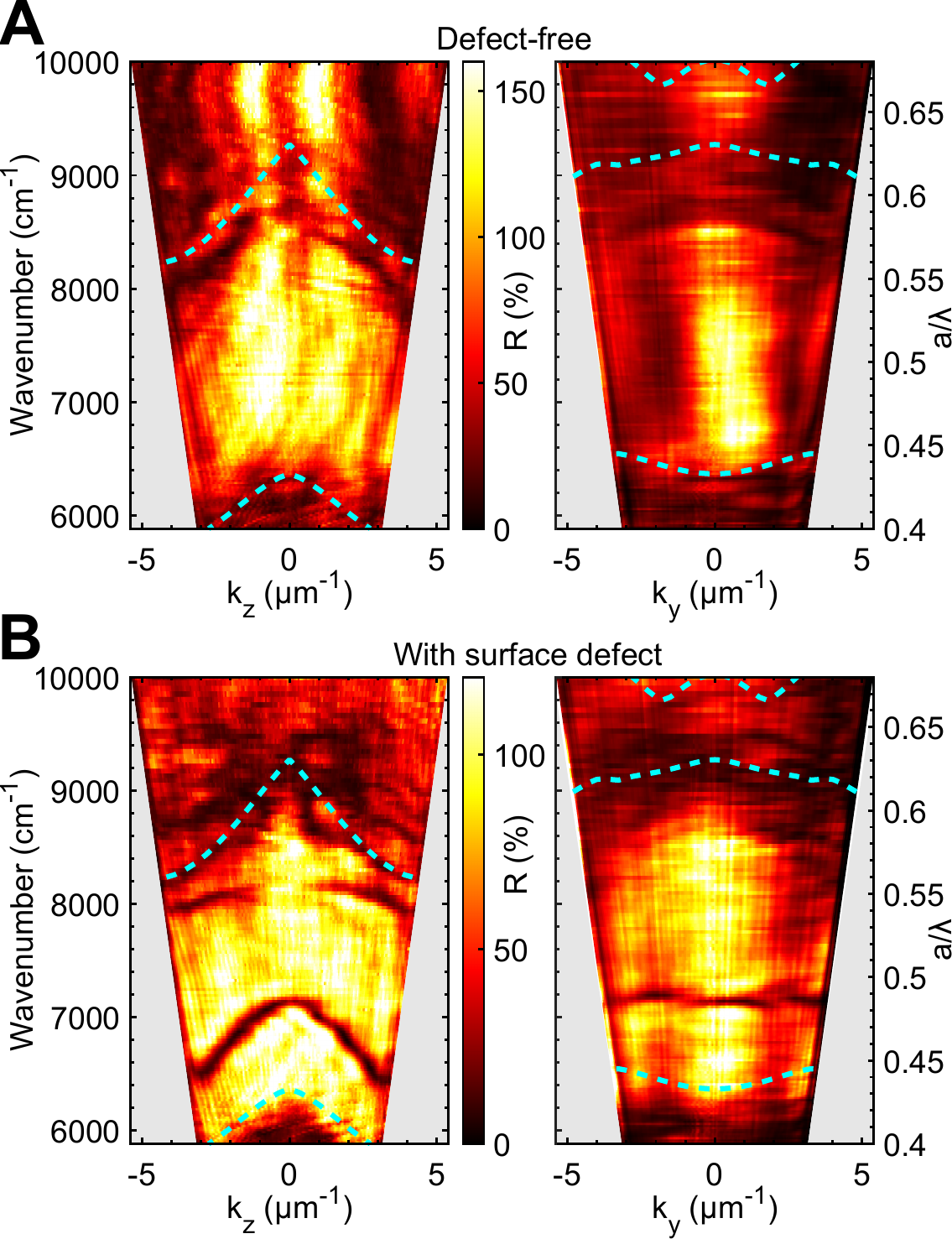}
    \caption{
    Momentum-resolved reflectivity $R$ for p-polarized light of a 3D photonic crystal (A) without and (B) with a surface defect.
    Samples without and with a surface defect are shown in Fig.~\ref{fig:sample3d}.
    (Gray area) Outside the objective's illumination cone.
    (Cyan, dashed) Symmetry-selected bands calculated along the red and blue lines in Fig.~\ref{fig:ibz_fbz} for pore diameter $d = 300$~nm. 
    }
    \label{fig:result3d_kykz_ppol}
\end{figure}

We now turn to the frequency- and momentum-resolved reflectivity spectra for p-polarized light.
To keep the representation of the large data sets tractable, we plot in Fig.~\ref{fig:result3d_kykz_ppol}(A) cross-sections through the momentum-resolved data at $k_y = 0$ (left) and $k_z = 0$ (right).
In the $k_y = 0$ plane, the reflectivity is high for all probed $k_z$ and symmetric about $k_z = 0$. 
With increasing frequency to about 8000~\pcm{}, low reflectivity appears at $\abs{k_z}$ = 3~\pum{} that we attribute to the upper edges of the band gap, also referred to as `conduction' bands~\cite{Joannopoulos2008book}.
Furthermore, decreased reflectivity appear at even higher frequencies, which we interpret as higher bands above the gap. 

In the $k_z$ = 0 plane in Fig.~\ref{fig:result3d_kykz_ppol}(A, right), we observe a high reflectivity for momenta central near $k_y = 0$, whereas the reflectivity decreases for $\abs{k_y} > 1$~\pum{}. 
These observations suggest that $\mathbf{G}_{\rm{dom}} = G_{\perp}\mathbf{\hat{x}}$ is valid in the $k_y = 0$ plane (left) but not in the $k_z = 0$ plane (right). 

Figure~\ref{fig:result3d_kykz_ppol} also shows (dashed cyan) bands calculated at the boundary of the first Brillouin zone, see the red and blue lines in Fig.~\ref{fig:ibz_fbz}.
The bands shown here are selected according to whether they can be excited with incident p-polarized light based on their dominant symmetry type.
Below the gap, the bands and reflectivity match well.
Above the gap, the shapes of the calculated modes match the shape of the measured band gap well, but the frequency of the calculated modes is greater than in the experiment for a straightforward reason: 
The lowest frequency of the first mode above the band gap occurs not at $\rm{S_0}$ but slightly inside the FBZ, as is seen by the frequencies of the modes calculated along $\Gamma$$\rm{M_0}$ in Fig.~\ref{fig:bandsGammaM0}.

\subsection{Frequency-resolved reflectivity of 3D band gap with surface defect}\label{sec:res_withSurfDef}
In Fig.~\ref{fig:result3d_kykz_ppol}(B), we plot the cross-sections of momentum-resolved reflectivity for a 3D photonic band gap crystal with a surface defect (see Figs.~\ref{fig:sample3d}(B,D)).
The left panel is a $k_y = 0$ cross-section, and the right panel a $k_z = 0$ cross-section.
Similarly to the crystal without a defect, the spectra are symmetric in $k_z = 0$ and $k_y = 0$, respectively. They are even more symmetric than without defect since fewer manufacturing steps are involved.

A striking feature in Fig.~\ref{fig:result3d_kykz_ppol}(B) is a deep trough inside the band gap. 
In the $k_y = 0$ cross-section (left), the trough runs from 6500~\pcm{} to 7050~\pcm{}, whereas the $k_z = 0$ cross-section (right) is nearly constant at 7050~\pcm{}. 
The width of the mode is about $\tilde{\nu} = 200$~\pcm{}, which corresponds to a relative bandwidth $\tilde{\nu}/\nu = 0.03$. 
The surface defect mode's dispersion is directly given by the shape of the troughs in these plots, and is very different between $k_z$ and $k_y$: $k_z$ strongly impacts the energy required to excite a surface defect mode, while $k_y$ hardly does. 
The group velocity is given by $\nabla_{\vb{k}} \omega$~\cite{Joannopoulos2008book}, where $\omega = 2\pi c_0 \tilde{\nu}$.
Given that the frequency of the surface defect mode decreases as a function of increasing $\abs{k_z}$, we expect the wave to travel in the opposite direction as $k_{\rm{in},\it{z}}$. 
Since the dispersion as a function of $k_y$ is relatively flat, we expect slow propagation in the $y$-direction. 

\subsection{Simulations of the surface defect mode}\label{sec:res_simulations}
\begin{figure}[t]
    \centering
    \includegraphics[width=\columnwidth]{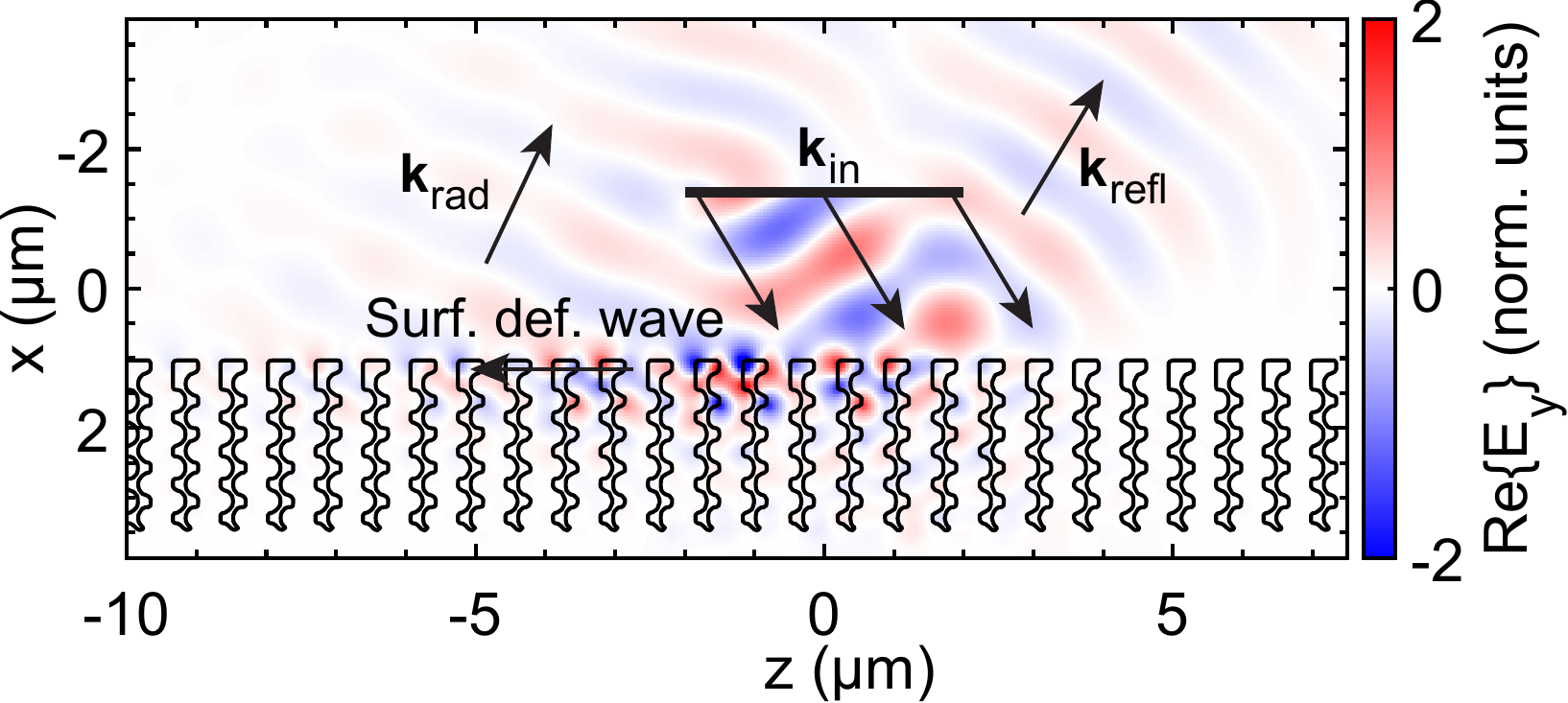}
    \caption{2D slice of a 3D FDTD simulation, with the real part of the electric field in the $y$-direction (Re$\{E_y\}$). 
    At the rectangular source (black horizontal bar), a continuous wave is created at the defect frequency and angle with wave vector $\kk{}_{\rm{in}}$. 
    The wave is incident on the photonic crystal with surface defect (black curves).
    The incident wave excites a surface defect wave in the \texttt{-}$z$-direction, which radiates in the direction of $\kk{}_{\rm{rad}}$.
    The incident wave is also partly reflected to $\kk{}_{\rm{refl}}$.
    } 
    \label{fig:simulationSurfWave30deg}
\end{figure}

Figure~\ref{fig:simulationSurfWave30deg} shows one snapshot of the FDTD simulations to visualize waves in the surface defect in real space, taken after 55 optical cycles.
The simulations show that a plane wave at \texttt{+}30\textdegree{} incidence is partly reflected towards $\kk_{\rm{refl}} \equiv \kk{}$, and partly excites a wave traveling between on one side the band gap structure, and on the other air.
From the decay of the E-field as a function of $x$, we estimate the penetration depth due to the band gap, also called the Bragg length, to be 700~$\pm 50$~nm, which corresponds to an effective permittivity of the band gap of about $\epsilon =$ \texttt{-}$0.12$.

The wave travels in the \texttt{-}$z$-direction, which is consistent with the measured dispersion in Sec.~\ref{sec:res_withSurfDef}.
It is astonishing that $k_{\rm{SDW},z}$ is negative (for negative $k_{\rm{in},z}$ like in Fig.~\ref{fig:illustrationSurfWave}) while the wave's energy travels to the right, \textit{i.e.}, the group velocity is positive~\cite{meepTutorial, Joannopoulos2008book}.
We conclude that we measure a backward propagating wave on the surface of a 3D photonic band gap crystal.
In the video in the Supplementary Material~\cite{supMat_c2_noDef}, we observe that the group velocity of the wave is slow compared to that of the reflected wave.

The wave radiates, and from the simulations we find it has a propagation length of $\sim3$~\um{}.
This is a useful start, especially since various improvements can be envisaged, see Sec.~\ref{sec:discussion}. 
The radiation travels in the \texttt{+}$z$-direction, in (almost) the same direction as $\kk$.
That is expected: in Sec.~\ref{sec:res_noSurfDef} we found that $k_{\rm{in},z} = k_{z}$, and due to conservation of parallel momentum $k_{\rm{in},z} = k_{\rm{SDW},z} = k_{\rm{rad},z}$, from which we conclude that $k_{z} = k_{\rm{rad},z}$.

The $\tilde{\nu}_{\rm{SDW}}$ that excites a surface defect mode at $k_z = 2.0$~\pum{} in the simulation is 270~\pcm{} smaller than in the experiment, which we attribute to the absence of manufacturing defects in~\cite{corbijn2023OptExp} and the limited resolution of the simulation.

While not shown here, in the $y$-direction, the wave hardly propagates near $\abs{k_y} = 0$, as expected from the measured dispersion along $k_y$.
Additional simulations show that at normal incidence ($k_y = k_z = 0$), the wave propagates in both $\pm z$-directions.

\subsection{Calculated dispersion via supercell method}\label{sec:res_dispCalculation}
\begin{figure}[t]
    \centering
    \includegraphics[width=\linewidth]{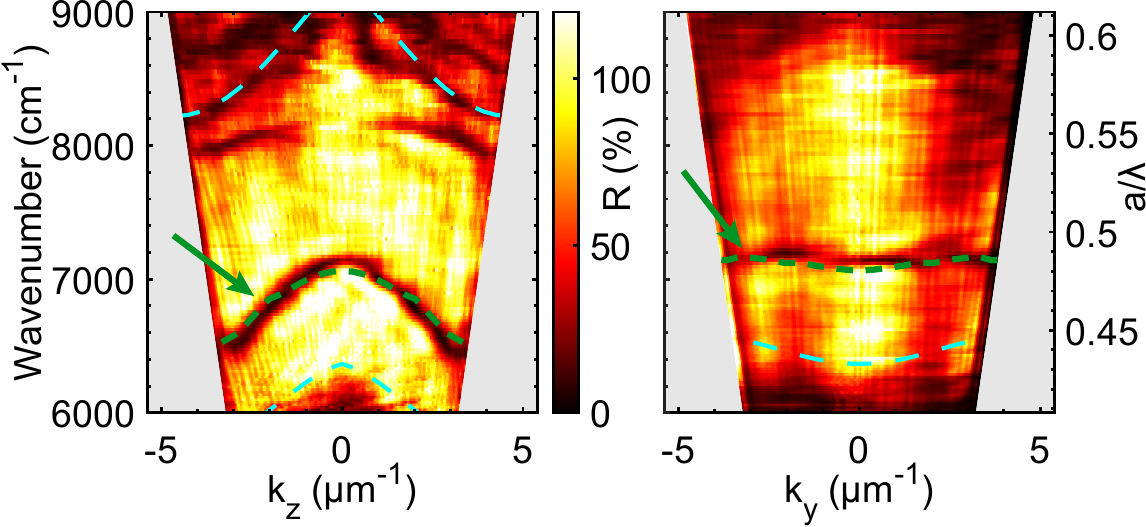}
    \caption{Zoom-in of Fig.~\ref{fig:result3d_kykz_ppol}(B) showing the momentum resolved reflectivity $R$ (colors in the color bar in the center) with (in green) the surface defect mode calculated with the supercell method.
    }
    \label{fig:defModeZoominWDispersion}
\end{figure}

Using the supercell method, we calculate the dispersion of the surface defect mode for the inverse woodpile with surface defect of $d_{\rm{def}} = 320$~nm.
Next to surface modes, the supercell method also provides bulk crystal and air modes, which are undesired here.
By setting requirements on the symmetry properties and confinement of the mode at the defect interface, we were able to isolate and reconstruct the dispersion of the defect mode, as further elaborated on in Appendix~\ref{sec:app_supercell}.
By again making the assumption that $\kk{}_{\rm{in},\parallel}$ = $\kk{}_{\parallel}$, which we argued is reasonable for a 2D surface defect in Sec.~\ref{sec:probe}, we can directly plot the (green dashed) calculated dispersion onto the measured data in Fig.~\ref{fig:defModeZoominWDispersion}.
The shape of the calculated dispersion matches very well with the measured dispersion, confirming the experimental results and theoretical analysis.
The defect mode is remarkably sensitive to the defect thickness $d_{\rm{def}}$: a change of $d_{\rm{def}}$ from 320 to 240~nm shifts the calculated defect frequency at $k_z = k_y = 0$ from 7060~\pcm{} to 7800~\pcm{} (Fig.~\ref{fig:mpb_SC_320nm}).

\section{Discussion}\label{sec:discussion}

\subsection{Didactic model}
The dispersion of the waves in the surface defect of the photonic crystal corresponds very well to the measured dispersion, but the good agreement in itself does not yet provide a physical argument as to why the waves are backward-propagating. 
To address this question, we propose an analytic model of Fresnel reflectivity of unstructured media, for p-polarized light. 
The model simplifies our experimental configuration by considering only three media, namely:\\

\noindent (i) an upper half-space consisting of air, with dielectric constant $\epsilon_1' = 1$,

\noindent (ii) a defect layer with a dielectric constant $\epsilon_2'$ and a thickness $d_{\rm{def}} = 320$~nm,

\noindent (iii) a lower half-space consisting of a photonic band gap crystal described by a negative real dielectric constant~\cite{Joannopoulos2008book, vinogradov2010surface, Hasan2018PRL} equal to $\epsilon_3' = -0.12$ \footnote{$\epsilon_3'$ was estimated from our FDTD simulations from the gradient of the electric field strength into the bulk of the band gap crystal (cf. Fig.~\ref{fig:simulationSurfWave30deg}). 
It turns out that the precise value of $\epsilon_3'$ is not critical, as long as $\epsilon_3' \ll \epsilon_2'$. }. \\

\noindent For the defect layer, we estimate the effective dielectric constant from a volume average of the refractive indices of air and silicon to be $\epsilon_2 = n_{\rm eff}^2 = 2.4^2$. 
The 3D photonic band gap is modeled with an effective dielectric constant that we take to be \textit{negative} and \textit{real}. 
The defect layer acts like a core in a low-$\epsilon$ cladding of a waveguide.
The (black) waveguide modes exist outside the ($m = 0$) light cone.

\begin{figure}
    \centering
    \includegraphics[width=0.8\linewidth]{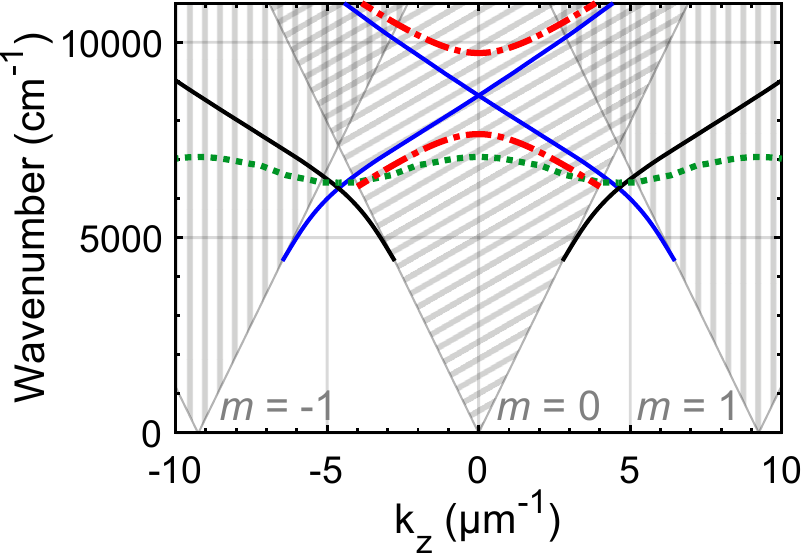}
    \caption{Calculated dispersion of modes in a Fresnel model of a much simplified sample of three slabs and a grating with orders $m = {-1, 0, 1}$. 
    Parameters of the model are defined in the text.
    (Striped fill) Light cones of $m = 0$ and $m = \pm 1$.
    (Solid lines) Calculated modes for (black) $m = 0$ and (blue) $m = \pm 1$.
    The calculated modes start from the edge of a light cone.
    (Green dotted) Supercell computations along $k_z$ from Fig.~\ref{fig:defModeZoominWDispersion}(left).
    (Red) Hyperbola originating from an avoided crossing of the blue $m = \pm 1$ modes.
    }
    \label{fig:disc_fresnelModel}
\end{figure}

Subsequently, we incorporate the effects of periodicity from the pores at the surface (illustrated in Fig.~\ref{fig:illustrationSurfWave}) using a grating with spacing $a$ such that $k_{\rm{in,z}}$ also creates cones at $m\frac{2\pi}{a} = m \cdot 9.24 $~\pum{} for $m \in \mathbb{Z}$, see the blue lines in Fig.~\ref{fig:disc_fresnelModel} for $m = \pm 1$.
The off-shell modes of $m = \pm 1$ represent backward waves below $\tilde{\nu} < 8600$~\pcm{} inside the light cone of $m = 0$, as can be seen from their negative dispersion.

In the model, the $m = \pm 1$ modes cross without interaction, but in reality, an avoided crossing is expected, which is taken into account by plotting a hyperbola with as asymptotes the $m = \pm 1$ modes, see the red line.
The opening $\Delta\tilde{\nu}$ at $k_z$ = 0 is estimated using Eq. 2.11 of Ref.~\cite{koenderink2003thesis} for a 2D structure with $\vb{G} = 4\pi/c \vb{\hat{z}}$.
Given the simplicity of the model, the hyperbola matches well with the supercell calculations at low $\abs{k_z}$. 
At high $\abs{k_z}$ the model's derivative does not tend to zero and thus the agreement with the supercell calculation is worse, which is reasonable as the effective medium approach of the defect layer is poorer for increasing $\abs{k_z}$.

Along the $y$-direction, the grating spacing is $a/\sqrt{2}$, which means the light cones are further apart and no grating effects are visible inside the frequency region of the band gap explored here.
Indeed, we do not observe a backward wave along the $y$-direction in our simulations.

\subsection{Further discussion}
Firstly, let us discuss the trade-off between ease of excitation of and amount of radiation by the surface defect wave. 
Using p-polarized waves inside the light cone, we excite p-polarized waves that radiate back into the air more readily than fully confined waves. 
For certain applications, less radiation and more confinement may be desired.
By increasing the confinement threshold to 80\% in the supercell computations, we find two strategies to improve the confinement:
(1) Confinement is enhanced by exciting (the same or other) bands at a $\sqrt{k_y^2+k_z^2}$ outside the light cone, \textit{e.g.}, by using a prism~\cite{Ishizaki2009nature,Saleh2019Wiley};
(2) by exciting surface modes with symmetry $\Sigma_1$ and $\Sigma_2$, which do not couple easily to outside waves.

Secondly, the question arises as to why we do not observe any defects in Fig.~\ref{fig:result3d_kykz_ppol}(A), where despite $d_{\rm{def}} = 0$ there is still a surface termination.
The reason is that there are no surface defect states in the band gap, which follows from the same calculations as in Sec.~\ref{sec:res_dispCalculation} with a confinement restriction at the bottom surface (which has defect width $d_{\rm{def}} = 0$).

Thirdly, at this time we have no simple explanation why there is such a strong focusing effect of incident light along the $y$-direction towards $k_y = 0$ for this sample in Fig.~\ref{fig:kspaceSingleFreq}(left). 
We do know that typically at least triple and sometimes quadruple multiple Bragg conditions are involved, that are complex in 3D photonic crystals. 

Fourthly, we discuss potential applications of a photonic crystal with such a surface defect.
Our hybrid system between a waveguide and a surface wave may not only have the usual ``flat light'' applications in (bio)sensing and photonic communication, but also completely new applications that profit from the 3D band gap (cavity quantum electrodynamics including spontaneous emission inhibition) combined with a tunable wave vector (hence: tunable directionality) that is set by the emission frequency. 
For example, say we put a quantum dot that emits between 6700 and 7100~\pcm{} inside our photonic crystal roughly 1.5~\um{} below the surface.
Due to the dispersion, when the quantum emitter emits in the surface defect light, the frequency sets the wave vector, hence the direction of the light. 
In other words, the emission frequency determines the direction: 
At 7100~\pcm{}, the angle with the $x$-axis at $k_y = 0$ is 0\textdegree{}, while at 6700~\pcm{} it is $\pm$28\textdegree{}.

\section{Conclusions and outlook}\label{sec:conclusion}
In this paper, we investigated the effect of a planar defect at the surface of a 3D photonic band gap crystal on its optical properties. 
Using momentum-resolved reflectivity spectra of p-polarized light, we probed the surface defect waves that show up as a deep minimum inside the high reflectivity of the 3D band gap. 
By tracking the troughs versus optical frequency and reflected wave vectors, we mapped out the dispersion relations of the defect states. 
To further interpret our experiments, we theoretically calculated the dispersion relations using a supercell method in the plane wave expansion, for which we selected modes based on symmetry and confinement properties expected for this Bloch mode inside a surface defect excited with p-polarized plane waves. 
The supercell calculations and experiments match very well and show a negative dispersion as a function of $\abs{k_z}$, indicative of a backward propagating wave, confirmed by finite-difference time-domain computer simulations. 

We anticipate that our experimental methods may be readily applied to two-dimensional (2D) photonic crystals, 2D photonic quasicrystals, other 3D photonic crystals, and other nanostructures, such as superlattices~\cite{Hack2019PRB}.
Momentum-resolved imaging provides a large dataset in a short amount of time, similar to \textit{e.g.} hyperspectral imaging, and it highlights the effect of defects very clearly.

\begin{acknowledgments}\label{sec:acknowledgments}
We thank Manashee Adhikary, Cornelis Harteveld, and Geert-Jan Kamphuis for their help with the optical setup and the 3D samples, Bert Mulder, Maël Hubert, and Devashish Sharma for discussions and help with the experiments, and William Wardley, Femius Koenderink, and Femi Ojambati for discussions on $\kk{}$-space imaging.
This work was supported by the project “Self-Assembled Icosahedral Photonic Quasicrystals with a Band Gap for Visible Light” (OCENW.GROOT.2019.071) of the “Nederlandse Organisatie voor Wetenschappelijk Onderzoek” (NWO); 
and by NWO-TTW Perspectief program P15-36 `Free-form scattering optics' (FFSO) with TUE, TUD, ASML, Demcon, Lumileds, Schott, Signify, and TNO; 
NWO-TTW Perspectief program P21-20 `Optical coherence; optimal delivery and positioning' (OPTIC) in collaboration with TUE, TUD, and ARCNL, Anteryon, ASML, Demcon, JMO, Signify, and TNO;
NWO JCSER Program, project 680-91-084, `Accurate and Efficient Computation of the Optical Properties of Nanostructures for Improved Photovoltaics';
and the MESA\texttt{+} Institute section Applied Nanophotonics (ANP). 
WLV wishes to dedicate this paper to the memory of Costas Soukoulis, one of the ``fathers'' of the inverse woodpile structure who greatly stimulated our research. 
\end{acknowledgments}

\appendix

\section{Crystal structure in real space}\label{sec:latticereal}
\begin{figure}[t]
    \centering
    \includegraphics[width=0.5\columnwidth]{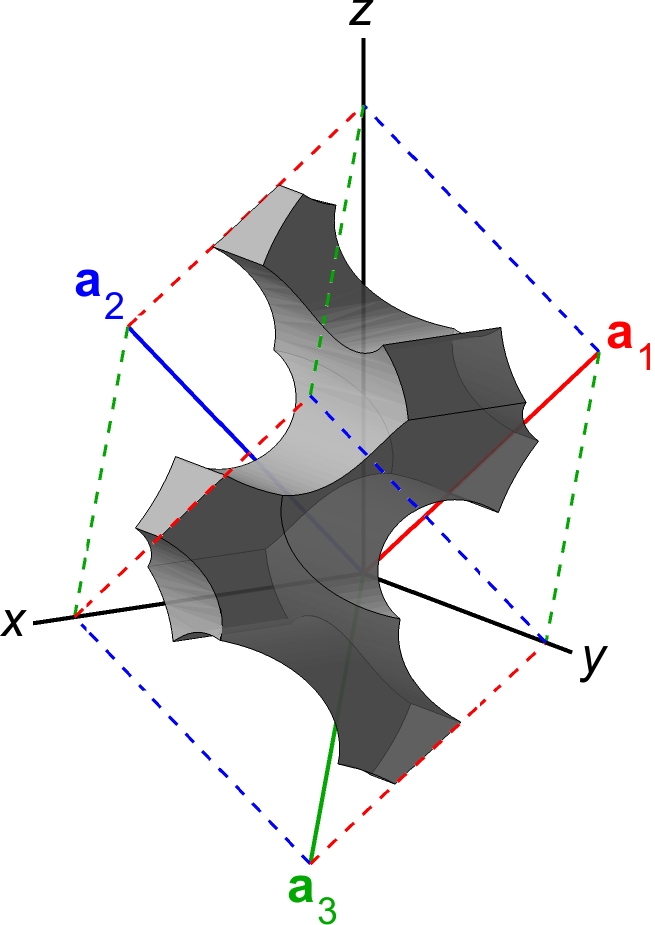}
    \caption{
    Unit cell of the inverse-woodpile structure. 
    In dark gray, the silicon structure is shown.
    The pore's material, air, is not shown here.
    The unit vectors and pores' positions are defined in the text.
    }
    \label{fig:unitcell}
\end{figure}

We wish to find the smallest unit cell describing the inverse woodpile structure.
In previous work~\cite{Woldering2009JAP, vandenBroek2012AFM, Devashish2017PRB, Hasan2018PRL, Adhikary2020OptEx,Woldering2009JAP, Huisman2011PRB}, an orthorhombic unit cell was used for convenience, but this unit cell is not primitive; 
hence, puzzling spurious artifacts occur in the band structures, such as the folding of bands. 
Instead, we use the primitive unit cell, more specifically, a body-centered tetragonal (BCT) unit cell with conventional unit vectors.
The unit cell is shown in Fig.~\ref{fig:unitcell}, and the unit vectors are given by
\begin{align}
\begin{split}
    \mathbf{a}_1 &= -\frac{a}{2\sqrt{2}} \hat{\mathbf{x}} + \frac{a}{2\sqrt{2}} \hat{\mathbf{y}} + \frac{a}{2} \hat{\mathbf{z}}, \\
    \mathbf{a}_2 &= \frac{a}{2\sqrt{2}} \hat{\mathbf{x}} - \frac{a}{2\sqrt{2}} \hat{\mathbf{y}} + \frac{a}{2} \hat{\mathbf{z}}, \\
    \mathbf{a}_3 &= \frac{a}{2\sqrt{2}} \hat{\mathbf{x}} + \frac{a}{2\sqrt{2}} \hat{\mathbf{y}} - \frac{a}{2} \hat{\mathbf{z}}.
\label{eq:unitVectors}
\end{split}
\end{align}

The structure has pores in the $x$- and $y$- directions.\footnote{To match with the symmetry tables, this notation differs from different work of our group where the pores were taken to be parallel to the $x$- and $z$-directions.} 
The space group of the structure is 141~\cite{hahn1983book}, see the Bilbao crystallographic server~\cite{aroyo2006bilbao2}.
To have the same basis for symmetry operations as the Bilbao server, we position the pore along the $x$-direction at $y = \frac{a}{4\sqrt{2}}$ and $z = \frac{a}{4}$, the pore along the $y$-direction at $x = z = 0$, and copy the pores at other positions inside the unit cell using the lattice vectors given in Eq.~\ref{eq:unitVectors}.


\section{Reciprocal lattice}
\begin{figure}[t]
    \centering
    \includegraphics[width=\columnwidth]{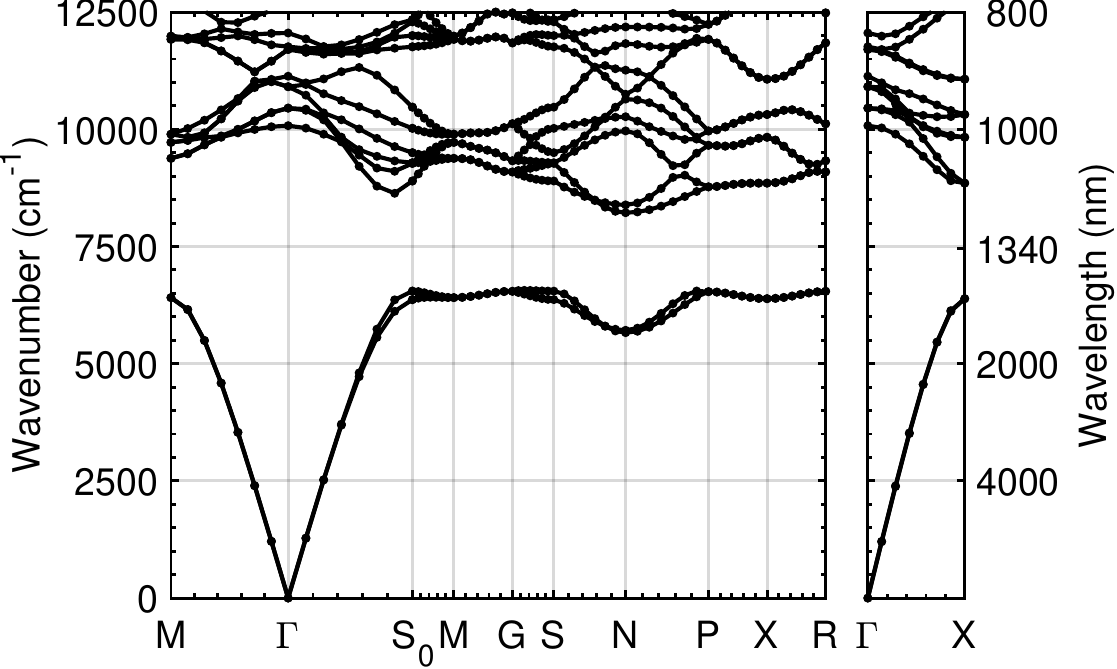}
    \caption{
    Band structure (dispersion diagram) of a 3D photonic crystal with an inverse-woodpile structure. 
    For this structure, the pore radius ($2r = d = 300$~$\si{\nano\meter}$) divided by the pitch ($a = 680$~$\si{\nano\meter}$) is equal to $r/a = 0.22$.
    }
    \label{fig:banddiagram}
\end{figure}

{\label{sec:latticereci}}
From the unit vectors, we obtain the reciprocal lattice vectors 
\begin{align}
\begin{split}
    \mathbf{b}_1 &= \frac{2\pi}{a}\left[\sqrt{2} \hat{\mathbf{y}} + \hat{\mathbf{z}}\right], \\
    \mathbf{b}_2 &= \frac{2\pi}{a}\left[\sqrt{2} \hat{\mathbf{x}} + \hat{\mathbf{z}}\right], \\
    \mathbf{b}_3 &= \frac{2\pi}{a}\left[\sqrt{2} \hat{\mathbf{x}} + \sqrt{2} \hat{\mathbf{y}}\right].
\end{split}
\label{eq:reciprocalVectors}
\end{align}
The commonly used irreducible Brillouin zone (IBZ) of the BCT lattice is shown in Fig.~\ref{fig:ibz_fbz}. 
The coordinates of the high-symmetry points of the IBZ are tabulated in Tab.~\ref{tab:symmetrycoordinates}. 
It is instructive to rotate the IBZ 180\textdegree{} around the NP line.
The rotated IBZ is one of the IBZs next to the non-rotated one, the unit cells together forming a triangular prism.
The procedure maps $\rm{S}_0$ to S, R to G, and M to $\rm{M_0}$, the latter positioned at $\frac{4}{3}\rm{S_0}$.
The band structure is shown in Fig.~\ref{fig:banddiagram}.

\begin{table}[t]
    \centering
    \setlength{\arrayrulewidth}{0.3mm}
    \setlength{\tabcolsep}{0pt}
    \renewcommand{\arraystretch}{1.2}
    \definecolor{mygray}{gray}{0.95}
    {\rowcolors{2}{mygray}{white}
    \begin{tabular}{ |>{\centering\arraybackslash}p{1.2cm}|>{\centering\arraybackslash}p{1.2cm}>{\centering\arraybackslash}p{1.2cm}>{\centering\arraybackslash}p{1.2cm}|>{\centering\arraybackslash}p{1.2cm}>{\centering\arraybackslash}p{1.2cm}>{\centering\arraybackslash}p{1.2cm}|  }
    \hline
    \textbf{Label} & $\mathbf{b_1}$ & $\mathbf{b_2}$ & $\mathbf{b_3}$ & $\textit{k}_\mathbf{x}$ & $\textit{k}_\mathbf{y}$ & $\textit{k}_\mathbf{z}$ \\
    \hline
    M       & 0.5    & 0.5   & -0.5   & 0          & 0          & 1     \\
    $\Gamma$  & 0      & 0     & 0      & 0          & 0          & 0     \\
    $\rm{S_0}$ & -0.375 & 0.375 & 0.375  & $\frac{3\sqrt{2}}{4}$ & 0    & 0     \\
    S       & 0.375  & 0.625 & -0.375 & $\frac{\sqrt{2}}{4}$   & 0    & 1     \\
    M       & 0.5    & 0.5   & -0.5   & 0          & 0          & 1     \\
    G       & 0.5    & 0.5   & -0.25  & $\frac{\sqrt{2}}{4}$   & $\frac{\sqrt{2}}{4}$ & 1 \\
    S       & 0.375  & 0.625 & -0.375 & $\frac{\sqrt{2}}{4}$   & 0    & 1     \\
    N       & 0      & 0.5   & 0      & $\frac{\sqrt{2}}{2}$   & 0    & 0.5   \\
    P       & 0.25   & 0.25  & 0.25   & $\frac{\sqrt{2}}{2}$   & $\frac{\sqrt{2}}{2}$ & 0.5 \\
    X       & 0      & 0     & 0.5    & $\frac{\sqrt{2}}{2}$   & $\frac{\sqrt{2}}{2}$ & 0     \\
    R       & -0.25  & 0.25  & 0.5    & $\frac{3\sqrt{2}}{4}$  & $\frac{\sqrt{2}}{4}$ & 0     \\
    $\Gamma$  & 0      & 0     & 0      & 0          & 0          & 0     \\
    X       & 0      & 0     & 0.5    & $\frac{\sqrt{2}}{2}$   & $\frac{\sqrt{2}}{2}$ & 0     \\
    \hline
\end{tabular}}
    \caption{Labels and corresponding positions along the irreducible Brillouin zone. 
    The first set of positions are in the units of the lattice vectors, Eq.~\ref{eq:reciprocalVectors}.
    The second set is coordinates in the Cartesian plane, for clarity divided by $2\pi/a$.
    The band structure is calculated along these points in this order.
    }
    \label{tab:symmetrycoordinates}
\end{table}

\section{Symmetry along the pores}{\label{sec:symalongpore}}
To excite a Bloch mode with a certain incident wave, their symmetry properties must match~\cite{Sakoda2005Springer}.
There are many different Bloch modes with various symmetry properties. 
Here, we restrict ourselves to waves with $\kk{}$-vectors along $\Gamma$$\rm{S_0M_0}$, \textit{i.e.}, plane waves in the $x$-direction.

As for the symmetry properties~\footnote{The symmetry properties of Bloch modes follow from the irreducible representations of the group of the wave vector. 
The given wave vectors are on the $\Sigma$ high symmetry line for which there are four irreducible representations: $\Sigma_1$ to $\Sigma_4$.} of the Bloch modes, there are four symmetry possibilities, $\Sigma_1$ to $\Sigma_4$, along the $\Gamma$$\rm{S_0M_0}$ line~\cite{aroyo2006bilbao2}.
As for the incident wave, s- and p-polarized plane waves are used in the lab.
We will elaborate on which symmetries ($\Sigma_1$ to $\Sigma_4$) we can excite using s- and p-polarized plane waves.

\begin{figure}[t]
    \centering
    \includegraphics[width=\columnwidth]{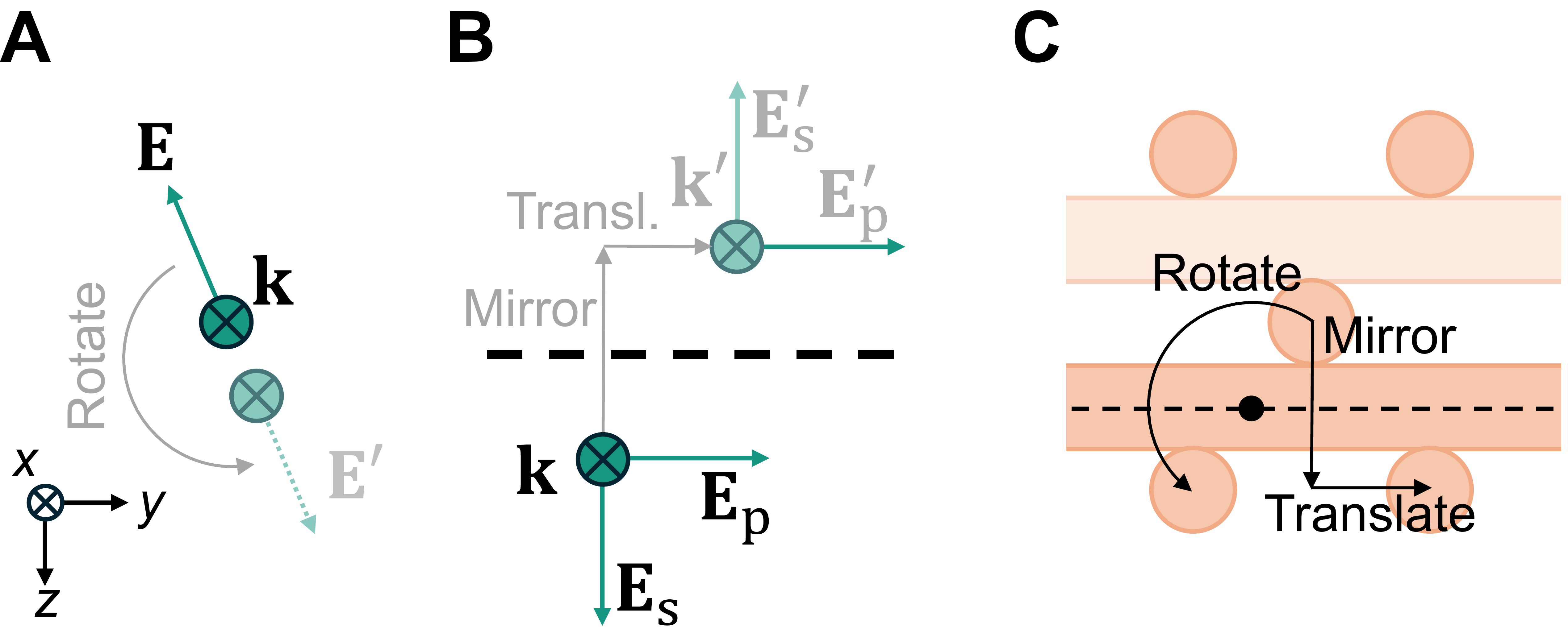}
    \caption{
    (A) A plane wave with $\rm{\mathbf{k}} = \it{k_x}\rm{\mathbf{\hat{x}}}$, and an electric field perpendicular to that.
    $\rm{\mathbf{E}}'$ is the result of the symmetry operation $\{2_{100}|0,0,0\}$ on $\rm{\mathbf{E}}$.
    (B) An s- and p-polarized plane wave with $\rm{\mathbf{k}} = \it{k_x}\rm{\mathbf{\hat{x}}}$. 
    $\rm{\mathbf{E}}'$ is the result of the symmetry operation $\{\rm{m}_{001}|0,\frac{1}{2},0\}$, which first mirrors in a $z$-plane, and then translates in the positive $y$-direction.
    (C) The crystal maps to itself under the symmetry operations $\{2_{100}|0,0,0\}$ and $\{\rm{m}_{001}|0,\frac{1}{2},0\}$. 
    The black dot denotes $x=y=z=0$. 
    }
    \label{fig:symSM}
\end{figure}

Three symmetry operations determine the symmetry of a Bloch mode along $\Gamma\rm{S_0M_0}$.
To start with, take the symmetry operation $\{2_{100}|0,0,0\}$, which rotates the electric field 180\textdegree{} around the $x$-axis.
From symmetry tables, we find that $\Sigma_1$ and $\Sigma_2$ states are symmetric for this operation, while $\Sigma_3$ and $\Sigma_4$ states are anti-symmetric.
Next, imagine a plane wave with $\rm{\mathbf{k}} = \it{k_x}\rm{\mathbf{\hat{x}}}$ and the $\EE{}$-field somewhere in the $yz$-plane, see Fig.~\ref{fig:symSM}(A).
The symmetry operation $\{2_{100}|0,0,0\}$ effectively flips the sign of the $\EE{}$-field.
Therefore, plane waves are always anti-symmetric for the $\{2_{100}|0,0,0\}$ operation, which means they can only excite $\Sigma_3$ or $\Sigma_4$ states.

The second symmetry operation of Bloch modes along the $\Gamma$$\rm{S_0M_0}$ line is $\{\rm{m}_{001}|0,\frac{1}{2},0\}$, which is a mirror operation in the $z$-plane, followed by a translation of half a unit cell in the $y$-direction. 
According to the symmetry tables, $\Sigma_3$ is anti-symmetric for this operation, while $\Sigma_4$ is symmetric.
Whether a plane wave is symmetric or anti-symmetric for this operation depends on the polarization of the plane wave.
Take a plane wave with $\rm{\mathbf{k}} = \it{k_x}\rm{\mathbf{\hat{x}}}$ and the $\rm{\mathbf{E}}$-field pointing in the \texttt{+}$z$-direction (s-polarized). 
Performing the operation $\{\rm{m}_{001}|0,\frac{1}{2},0\}$ on this wave results in a $\rm{\mathbf{E}}$-field that is pointing in the \texttt{-}$z$-direction, see Fig.~\ref{fig:symSM}(B), and thus the s-polarized wave is anti-symmetric for this symmetry operation, just like $\Sigma_3$. 
On the contrary, a p-polarized wave has an $\rm{\mathbf{E}}$-field in the $y$-direction, which is symmetric for this symmetry operation, just like $\Sigma_4$.
Note that the translation does not matter for the plane wave because $\kk{} = k_x \vb{\hat{x}}$ is orthogonal to the translation.

The third symmetry operation, $\{\rm{m}_{010}|0,\frac{1}{2},0\}$, does not provide additional constraints, and is therefore not detailed further.

\begin{figure}[t]
    \centering
    \includegraphics[width=0.7\columnwidth]{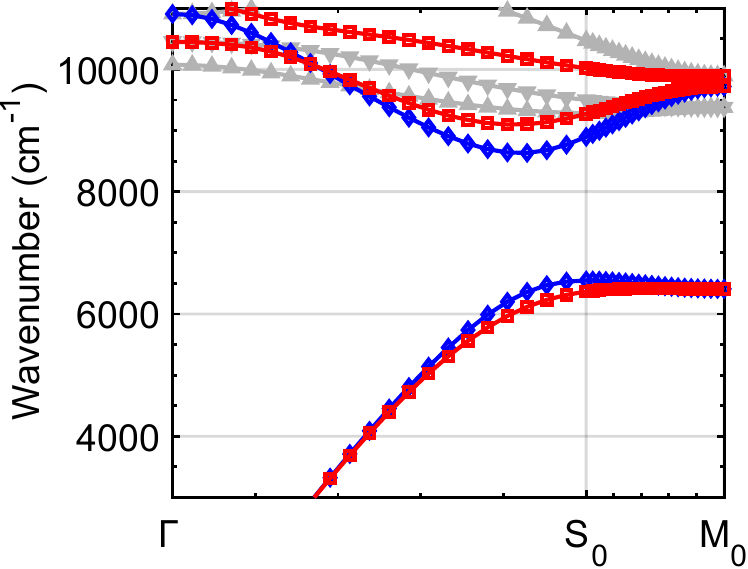}
    \caption{
    Bands calculated along $\Gamma\rm{S_0M_0}$ with symmetry properties (blue diamonds) $\Sigma_3$, (red squares) $\Sigma_4$, (grey downwards triangles) $\Sigma_1$, and (grey upwards triangles) $\Sigma_2$.
    Calculated for pore diameter 300~nm.
    s-Polarized waves only excite $\Sigma_3$ modes, and p-polarized waves $\Sigma_4$ modes.
    }
    \label{fig:bandsGammaM0}
\end{figure}

As an example, in Fig.~\ref{fig:symSM}(C) the first and second symmetry operations are performed on one point, 
namely $x = 0$, $y = a/(4\sqrt{2})$, and $z = a/4$.

In summary, along the $x$-direction, s-polarized (p-polarized) plane waves only excite modes in the $\Sigma_3$ ($\Sigma_4$) symmetry group.
Besides, modes in the $\Sigma_1$ and $\Sigma_2$ groups cannot be excited with plane waves due to symmetry constraints.
The bands along the $\Gamma$$\rm{S_0M_0}$ path are shown with their symmetry properties in Fig.~\ref{fig:bandsGammaM0}.
\section{Supercell calculation results}{\label{sec:app_supercell}}
Using the supercell method, we calculate the frequency of the surface defect mode as a function of $k_y$ and $k_z$.
However, we also obtain the frequency of the air and photonic crystal modes, which we wish to remove.
The strategy of Ref.~\cite{meade1991prb} to keep only the modes that do not shift when adjusting the number of cells requires many calculations and does not provide enough information here, so we use a different approach based on confinement and symmetry, which we will explain.

A surface mode represents energy confined near the surface.
Therefore, we calculate the energy density of each mode, and consider a mode to be confined near the surface if at least 50\% of the mode's energy density is located within one unit cell about the surface.
Additionally, as we only experimentally probe the crystal with p-polarized light, the symmetries of fields of the calculated modes are also important.
The symmetry types are only defined for waves with exactly $\kk{} = k_x \vb{\hat{x}}$, but we define a \textit{most-dominant} symmetry type for off-axis modes if their field is at least 50\% symmetric or anti-symmetric for all three symmetry operations in Sec.~\ref{sec:symalongpore}.

\begin{figure}[t]
    \centering
    \includegraphics[width=\linewidth]{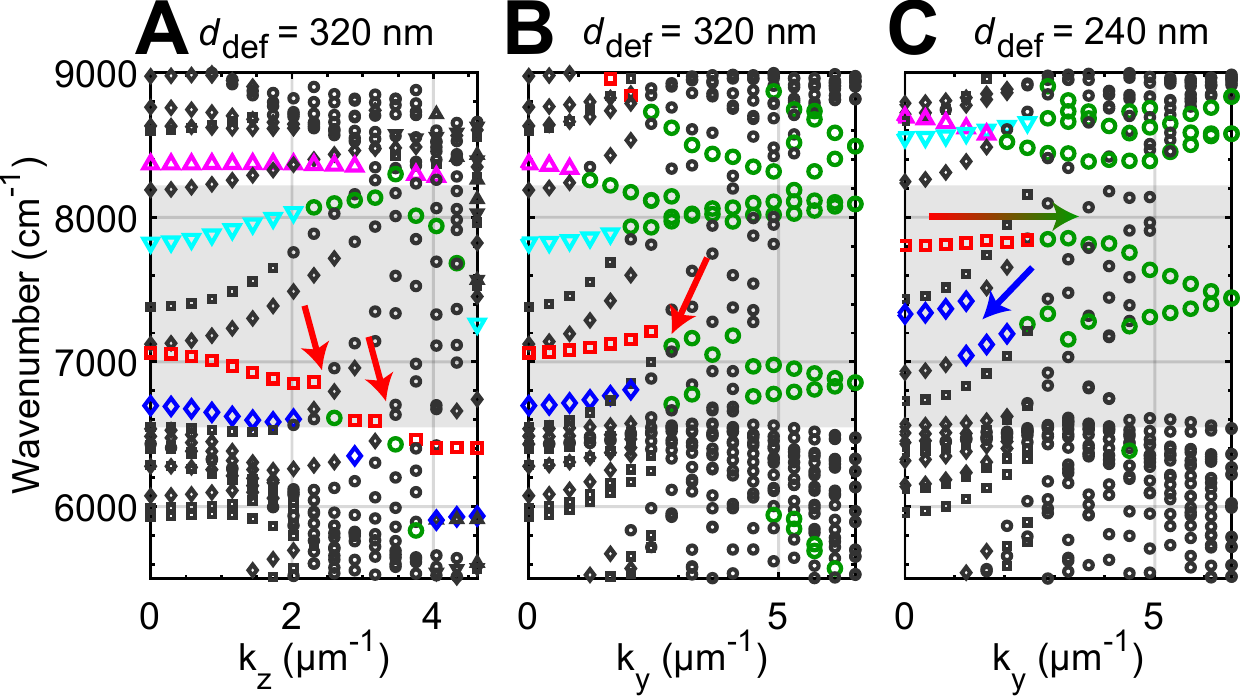}
    \caption{Bloch modes calculated along the $k_z$ and $k_y$-directions using the supercell method.
    (Black) Modes not confined at the surface.
    (Color) Modes confined at the surface.
    Downwards triangles, upwards triangles, diamonds and squares correspond to $\Sigma_1$ to $\Sigma_4$, respectively.
    Circles have no categorized symmetry type.
    (Gray) Calculated band gap.
    (Red/blue arrows) Examples of points where the air modes interfere with the surface defect mode.
    (Red to green arrow) The polarization becomes less p-polarized, eventually below the threshold defined in the text.}
    \label{fig:mpb_SC_320nm}
\end{figure}

The Bloch modes from the supercell method calculated for a structure with $d_{\rm{def}} = 320$~nm are shown in Fig.~\ref{fig:mpb_SC_320nm}(A,B), where the red squares pertain to predominantly p-polarized surface defect modes, \textit{i.e.}, the modes we can excite in our experiment.
The calculated dispersion as a function of $k_z$ is mostly smooth, but there is disturbance by the air modes between $k_z \in [2.31, 2.60]$~\pum{} and between $k_z \in [3.18,3.46]$~\pum{}, see the red arrows in Fig.~\ref{fig:mpb_SC_320nm}(A).
We therefore disregard the states in those regions to obtain a good estimate of the dispersion as a function of $k_z$. 

The disturbance is much stronger against $k_y$, where the surface mode is pushed up by a p-polarized air mode near $k_y$ = 2~\pum{} at the red arrow in Fig.~\ref{fig:mpb_SC_320nm}(B).
Near the same point, the mode becomes less p-polarized, eventually below the threshold defined earlier, \textit{i.e.}, it becomes uncategorized (green circles).
As there are no confined modes between 7200~\pcm{} $< \tilde{\nu} <$ 7800~\pcm{}, we know that the mode should follow the uncategorized mode to $\tilde{\nu} =$ 6860~\pcm{}.
The calculated p-polarized modes between 1~\pum{} $< k_y < $ 5~\pum{} do not accurately describe the dispersion of the surface defect mode.
\footnote{To provide an additional example, the effect of the disturbance of air modes with a surface mode is clearer for the s-polarized modes (blue diamonds).
The hybridization of the s-polarized mode defect mode with the air mode is especially visible for $d_{\rm{def}} = 240$~nm near $k_y = 1$~\pum{} (blue arrow).}
To obtain the shape of the mode, the same calculation is performed with $d_{\rm{def}}$ = 240~nm, where the p-polarized defect mode is hardly disturbed by air modes, see Fig.~\ref{fig:mpb_SC_320nm}(C).
Besides, the red squared data points clearly continue in the green circular data points.
To obtain a good estimate of the dispersion of the surface mode as a function of $k_y$, we linearly scale the dispersion of the mode of $d_{\rm{def}} = 240$~nm to the beginning and endpoint of the mode of $d_{\rm{def}} = 320$~nm.
The resulting dispersion as a function of $k_z$ and $k_y$ are shown in Fig.~\ref{fig:defModeZoominWDispersion}.

\section{List of measurements}{\label{sec:app_measuredsamples}}
In our study, we collected momentum-resolved reflectivity on two different structures, which were etched simultaneously on the same beam.
All measurements are listed in Tab.~\ref{tab:samplesMeasured}.
We only show the first and second sets of measurements in this document, and of those, except for Fig.~\ref{fig:realreflSpectrum}, only p-polarized light for the following reasons:
(1) Crystal number~3 has smaller pores by design, leading to a band gap 100~\pcm{} lower than that of crystal~2 for p-polarized light, which is closer to the lower edge of our detection regime.
The dispersion of crystal 3's surface defect has the same shape as crystal 2's, as is observed in the additional Supplementary Video~\cite{supMat_c2_noDef}.
(2) Waves incident from the $x$-direction ($y$-direction) are incident perpendicular to the first (second) etch direction.
As the first etch step is less complex, the dispersion of the surface defect mode is more symmetric for momentum-resolved imaging from the $x$-direction than the $y$-direction.
(3) The s-polarized surface defect mode (blue diamonds in Fig.~\ref{fig:mpb_SC_320nm}) is not as clear in experiment as the p-polarized mode, probably because the frequency of the s-polarized mode is almost at the bottom of the band gap.

\begin{table}[t]
    \centering
    \setlength{\arrayrulewidth}{0.3mm}
    \setlength{\tabcolsep}{0pt}
    \renewcommand{\arraystretch}{1.2}
    \definecolor{mygray}{gray}{0.95}
    {\rowcolors{2}{mygray}{white}
    \begin{tabular}{ |>{\centering\arraybackslash}p{1cm}>{\centering\arraybackslash}p{1.7cm}>{\centering\arraybackslash}p{1.7cm}>
    {\centering\arraybackslash}p{1.7cm}>{\centering\arraybackslash}p{1.7cm}|}
    \hline
    \textbf{Beam} & \textbf{Crystal number} & \textbf{Incidence axis} & \textbf{Surface defect} & \textbf{Maximum R~(\%)} \\
    \hline
    15F  & 2  & $x$   & present & 75 \\
    15F  & 2  & $x$   & absent & 80  \\
    15F  & 2  & $y$   & present & 60 \\
    15F  & 2  & $y$   & absent & 64  \\
    15F  & 3  & $x$   & present & 79 \\
    15F  & 3  & $x$   & absent & 75  \\
    15F  & 3  & $y$   & present & 65 \\
    15F  & 3  & $y$   & absent & 68  \\
    \hline
\end{tabular}}
    \caption{List of measurements performed for this paper. 
    }
    \label{tab:samplesMeasured}
\end{table}

\bibliography{references} 
\end{document}